\documentclass[12pt]{iopart}
\usepackage{graphicx}
\usepackage{epsfig}
\expandafter\let\csname equation*\endcsname\relax
\expandafter\let\csname endequation*\endcsname\relax
\usepackage{amsmath}
\usepackage{iopams}

\newcommand{\bra}[1]{\ensuremath{\langle{#1}|\,}}
\newcommand{\ket}[1]{\ensuremath{\,|{#1}\rangle}}
\newcommand{\braket}[1]{\ensuremath{\langle{#1}\rangle}}

\begin{document}

\title
{Nonequilibrium configuration interaction method for transport in  correlated quantum systems}

\author{Alan A. Dzhioev$^1$ and D. S. Kosov$^2$}

\address{$^1$ Bogoliubov Laboratory of Theoretical Physics, Joint Institute for Nuclear Research,  RU-141980 Dubna, Russia}

\address{$^2$ School of Engineering and Physical Sciences, James Cook University, Townsville, QLD, 4811, Australia }

\pacs{05.60.Gg, 73.63.-b, 72.10.Di, 05.70.Ln, 31.15.V-}

\begin{abstract}
We present a new approach to treat correlations in nonequilibrium quantum many-particle system.
The method is based on ideas of  configuration interaction theory of exact nonperturbative ground state electronic structure calculations. We use superoperator techniques in  Liouville-Fock space and   represent the  nonequilibrium density
matrix  as a linear combination of all possible nonequilibrium quasiparticle excitations built on the
appropriate reference state. As an example we consider the electron transport through the system with electron-phonon interaction.
The concept of embedding (buffer zones between the reservoirs and the correlated quantum system) is  used to derive an exact master equation for the reduced density matrix.
Using approximate (truncated) expansion of the trial density matrix we obtain the linear system of equations for two-quasiparticle amplitudes.
Then we compute the steady-state
current and compare the result with other approaches. The current conserving property of the method is proved.
\end{abstract}

\date{\today}
\maketitle

\section{Introduction}

The description of correlated quantum many-body systems  far from equilibrium remains one of the challenging problems in modern statistical mechanics, quantum field theory, nuclear, atomic, molecular
and  condensed matter physics.\cite{szymanska,berges-review,rammer-book} Most  of theoretical approaches are based on Keldysh nonequilibrium Green's functions (NEGF) which enables  perturbative
treatment of nonequilibrium many-body systems.\cite{keldysh65,rammer-book}
NEGF also allows systematic summation of specific classes of nonequilibrium diagrams, for example, random phase approximation\cite{PhysRevB.80.165305} or GW theory.\cite{thygesen07}
Nonperturbative methods, such as, numerical renormalization group theory, are also available but often restricted to the
oversimplified model systems.\cite{nrg12} In this paper we propose a new theoretical method -- nonequilibrium configuration interaction (NECI) which,
in principle, is able to achieve the exact solution of the nonequilibrium problem.
The suggested approach is  not based on the perturbative expansion in the interaction strength (electron-electron or electron-phonon) but provides an expansion
of the nonequilibrium density matrix in terms of multi-quasiparticle and multi-phonon excitations over the reference state. Each term in such
expansion includes an infinite series in the interaction strength and for this reason we consider our method as an nonperturbative one.

The remainder of the paper is organized as follows. In Sec.~2, we introduce the Lindblad master equation for an embedded system
and its superoperator representation. Section~3 presents the nonequilibrium configuration interaction method. We
discuss two different truncated expansions for the density matrix and apply the theory to calculate the steady-state current.
In this section, we also proof that NECI method is current conserving.
The numerical NECI calculations of the steady-state current  and comparison with the results obtained with the nonequilibrium Green's function formalism are presented in Sec.~4.
Conclusions are given in Sec.~5. Appendix~A contains a summary of the superoperator formalism.
We use natural units throughout the paper: $\hbar= k_B = |e| = 1$, where $-|e|$ is the electron charge.

\section{Lindblad master equation in  Liouville-Fock space}

\begin{figure}[t!]
\begin{center}
\includegraphics[width=0.7\columnwidth]{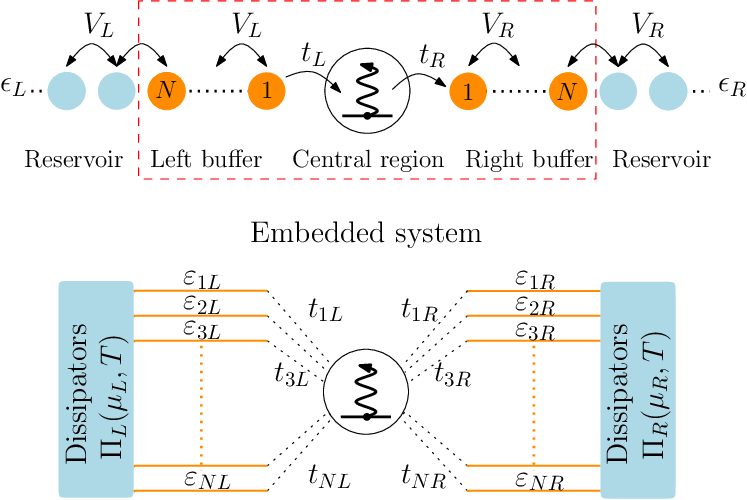}
\end{center}
\caption{Schematic illustration of embedding of an open, interacting quantum system. Upper part: Each electrode is divided into a macroscopic reservoir  and a finite buffer zone.
The region enclosed by the red dashed line is the embedded system. Lower part: The projection of the reservoirs  results into the Lindblad master equation for the reduced density matrix of the embedded quantum system. Each buffer zone energy level $\varepsilon_{k\alpha}$ is connected to the dissipator and the central region. }
\label{system}
\end{figure}

As shown in Figure~\ref{system}, we consider
a model of nonequilibrium many-body theory: the interacting  quantum system (central region) connected to two  noninteracting electrodes, left ($L$) and right ($R$),
maintained  with different chemical potentials, $\mu_L$ and $\mu_R$, and the same  temperature $T$.
To be specific, we focus on the electron transport problem through
one electronic single-particle level with energy $\varepsilon_0$ coupled linearly to a vibrational mode (phonon) of frequency $\omega_0$
(the so called local Holstein model). Thus, the central region Hamiltonian is
\begin{equation}\label{H_lHm}
  H_S = \varepsilon_0 a^\dag a + \omega_0 d^\dag d + \kappa a^\dag a(d^\dag + d),
\end{equation}
where $a^\dag$ ($a$) and  $d^\dag~(d)$ are electron and phonon creation (annihilation) operators, respectively. The left and right electrodes
are modeled as semi-infinite tight-binding  chains of atoms characterized by the hopping matrix element $V_\alpha$ and the on-site energy $\epsilon_\alpha$ ($\alpha=L,~R$). The coupling between the cental region and the end site of $\alpha$ electrode is given by the matrix element $t_\alpha$.

We replace the infinite system by a finite one using open boundary conditions. To this end
each electrode is partitioned into two parts: the macroscopically large reservoir and the finite buffer zone between the central region and the reservoir.
\footnote{The idea of the buffer zone was first proposed by us\cite{dzhioev11a,dzhioev11b,dzhioev12} and then also employed  by Ajisaka et al.\cite{prosen12}}
  We assume that each buffer zone contains $N$ atoms. The Hamiltonian of the whole system takes the  form
\begin{equation}
{\cal H} = H_{S}+ H_{SB} + H_B + H_{BR} + H_R.
\label{h}
\end{equation}
In the energy representation  the reservoir and the buffer zone Hamiltonians are diagonal
\begin{equation}
{  H}_R = \sum_{  l \alpha  }  \varepsilon_{l\alpha } a^{\dagger}_{ l \alpha  } a_{ l \alpha },~~
{  H}_B = \sum_{ k \alpha}  \varepsilon_{ k\alpha } a^{\dagger}_{k \alpha  }  a_{k \alpha },
\end{equation}
where $\varepsilon_{l\alpha }$ denote the continuum single-particle spectra of the left ($\alpha=L$) and right ($\alpha=R$) reservoir states,
while the buffer zones have discrete energy spectrum $\varepsilon_{k\alpha }$ ( $k=1,\ldots, N$).
The buffer-reservoir and central region-buffer couplings  have the standard tunneling form:
\begin{equation}
{  H}_{BR} = \sum_{ lk \alpha}  ( v_{lk \alpha } a^{\dagger}_{l \alpha   }  a_{k \alpha  } + \mathrm{h.c.}) ,
\label{v}
\end{equation}
\begin{equation}
{H}_{SB} = \sum_{ k \alpha  }  ( t_{k \alpha } a^{\dagger}_{k \alpha } a + \mathrm{h.c.} ).
\label{t}
\end{equation}

Projecting out the reservoir degrees of freedom from the Liouville-von Neumann equation for the total density matrix, we obtain
Lindblad master equation for the reduced density matrix of the  embedded system (central region + finite buffer zones)
\cite{dzhioev12, stability}
\begin{align}\label{lindblad}
& i \frac{\partial\rho(t)}{\partial t} =[H,\rho(t)] + i{\Pi}\rho(t).
\end{align}
Here, $H$  is the Hamiltonian of the embedded system which includes the Lamb shift of the buffer zone single-particle levels
\begin{equation}\label{embedded_H}
H= H_S +H_{SB} +H_B +\sum_{k \alpha }\Delta_{k\alpha}   a^{\dagger}_{k \alpha} a_{k \alpha},
\end{equation}
and ${\Pi}\rho(t)$ is the non-Hermitian dissipator  given by the standard Lindblad form
\begin{align}\label{non_herm}
{\Pi}\rho(t) = \sum_{k\alpha} \sum_{\mu=1,2} \bigl(2 L_{ k\alpha   \mu}\rho(t) L^\dag_{ k\alpha  \mu} - \{L^\dag_{k \alpha \mu} L_{ k \alpha  \mu},\rho(t) \}\bigr).
\end{align}
The operators $L_{k\alpha 1}$ and $L_{k\alpha 2}$ are referred to as the Lindblad  operators, which represent the buffer-reservoir interaction. They have the following
form:
\begin{align}\label{L_operators}
  L_{k\alpha 1} = \sqrt{\Gamma_{k\alpha 1}}a_{k \alpha},~~L_{k \alpha  2} = \sqrt{\Gamma_{k\alpha 2}  }a^\dag_{ k  \alpha}.
\end{align}
with $\Gamma_{k\alpha 1}=\gamma_{k\alpha}(1-f_{k\alpha})$,  $\Gamma_{k\alpha 2}=\gamma_{k\alpha}f_{k\alpha}$. Here $ f_{k\alpha} =[1+ e^{(\varepsilon_{k\alpha} - \mu_\alpha)/T }]^{-1}$
\emph{}and   $\gamma_{k\alpha} $ ($\Delta_{k\alpha}$) is the imaginary (real)  part of the self-energy arising from the buffer-reservoir interaction
$\sum_l |v_{lk\alpha }|^2/(\varepsilon_{k\alpha}-\varepsilon_{l\alpha } +i0^+)$.

The Lindblad master equation~\eqref{lindblad} describes the time evolution of an open  embedded quantum system preserving   Hermiticity, normalization, and positivity of the nonequilibrium density
matrix. Open boundary conditions are taken into account by the non-Hermitian dissipative part, $\Pi \rho(t)$, which represents the
influence of the reservoir on the buffer zone.  The applied bias potential enters into the master equation
via fermionic occupation numbers $f_{k\alpha}$ which depend on the temperature and the chemical potential in the left
and right electrodes.
We have recently demonstrated that this embedding procedure makes the master equation (\ref{lindblad}) exact in the steady-state regime, if the buffer zones are large enough to cure the deficiencies of Born-Markov approximation for treating the buffer-reservoir interface.\cite{dzhioev11a,dzhioev11b,dzhioev12}

The Lindblad master equation ~\eqref{lindblad} is not gauge invariant with respect to the applied voltage bias, which is the difference between the chemical potentials of left and right reservoirs.
To make the theory gauge invariant (for static electric field) one should add electrostatic potential which is obtained by solving Poisson equation with appropriate boundary conditions (electrictrostatic potential should be $\mu_L$ at the left boundary and $\mu_R$ at the right boundary).  We do not solve the Poisson equation in this paper.  Since we anyway choose  the model hamiltonian for the buffer zones and for the system, the introducing the Poisson equation is unnecessary complication which does not have any importance for the results discussed  here.

We use a superoperator formalism and  convert the  Lindblad master equation~\eqref{lindblad} to the non-Hermitian Schr\"{o}edinger-like equation for the nonequilibrium density matrix.\cite{dzhioev11b,dzhioev12} Within the superoperator formalism the density matrix is considered as a vector in the Liouville-Fock space and all other operators as superoperators.
In fact this Liouville-Fock space is nothing else but the CAR/CCR algebra of observables which admit
the Gelfand-Naimark-Segal (GNS) representation, while the Liouville superoperator defined
below is the generator of dynamics in this larger space.
 In Appendix A we present a brief summary of the formalism and demonstrate how to convert the Lindblad master equation into a superoperator form.
The resulting equation is
\begin{equation}\label{Schrodinger}
   i\frac{\partial}{\partial t}\ket{\rho(t)} = (\hat H - \widetilde H -i\sum_{k\alpha}\Pi_{k\alpha})\ket{\rho(t)}=L\ket{\rho(t)}.
\end{equation}
Here, the superoperators $\hat H = \hat H_S + \hat H_{SB} + \hat H_B$ and $\widetilde H = \widetilde H_S + \widetilde H_{SB} + \widetilde H_B$ are obtained from the Hamiltonian~\eqref{embedded_H} by replacing ordinary creation and annihilation operators $a^\dag,~a$ with non-tilde, $\hat a^\dag,~\hat a$, and tilde, $\widetilde a^\dag,~\widetilde a$, superoperators, respectively (note that we include the Lamb shift into $H_B$). The dissipators are given by
\begin{align} \label{Pi}
\Pi_{k\alpha} =&(\Gamma_{k\alpha1}-\Gamma_{k\alpha2})(\hat a^\dag_{k\alpha}\hat a_{k\alpha}+
  \widetilde a^\dag_{k\alpha}\widetilde a_{k\alpha})
  \notag\\&-
  2i(\Gamma_{k\alpha1}\widetilde a_{k\alpha}\hat a_{k \alpha}+\Gamma_{k\alpha2}\widetilde a^\dag_{k\alpha}\hat a^\dag_{k\alpha})+2\Gamma_{k\alpha2}.
   \end{align}
The Liouville superoperator $L$ is non-Hermitian because of $\Pi_{k\alpha}$. Some other important properties of $L$ are discussed in Appendix A. In particular, the relation $\bra{I}L=0$ holds,
where the Liouville-Fock space bra-vector $\bra{I}$ corresponds to the unit operator~$I$.

We focus our attention on asymptotic ($t\to\infty$) nonequilibrium steady-state situation, where the density matrix $\ket{\rho(t)}$ does not depend on time.
Therefore the problem is reduced to the problem of finding the  right zero-eigenvalue eigenvector (i.e., the null-space) of the non-Hermitian infinite-dimensional  Liouville superoperator
\begin{equation}\label{st-st_eq}
  L\ket{\rho}=0.
\end{equation}
Knowing the solution of~\eqref{st-st_eq}, we can compute the steady-state current into the central region from $\alpha$ buffer zone as
\begin{equation}\label{exp_value}
 J_\alpha = \braket{I|\hat J_\alpha|\rho},
\end{equation}
where
\begin{equation}
  \hat J_\alpha = -i\sum_{k}t_{k\alpha}(\hat a^\dag_{k\alpha}\hat a - \hat a^\dag\hat a_{k\alpha})
\end{equation}
is the current superoperator.

In the next section we discuss the development  of NECI method to solve equation~\eqref{st-st_eq}.
We also present the application of NECI  to compute the steady-state current through the central region. But
before proceeding, we would like to say some   words about the existence and uniqueness of the solution of Eq.~\eqref{st-st_eq}.
For a finite open quantum system described by the Lindblad master equation
the sufficient conditions for the existence and uniqueness of the steady
state are provided by the Spohn theorems~\cite{Spohn1976,Spohn1977}. To the best of our knowledge, there is no
explicit conditions guaranteeing the uniqueness  of the solution of Eq.~\eqref{st-st_eq} for infinite systems like the considered model with electron-phonon
correlations. However, if we introduce the physical
high-energy cut off for phonon states we find that the conditions of Spohn's theorems are satisfied. This gives us the good reason to think (and this observation is also supported by our numerical calculations)  that the solution of Eq.~\eqref{st-st_eq} exists and it is unique.

\section{Configuration interaction method for nonequilibrium density matrix}

In the beginning of the section, we
make the important remark on the notation use in the rest of the paper: only creation/annihilation
superoperators written with letters $a,~d$ (such as for example $\hat a_{b \alpha}$ and $\hat a^{\dagger}_{b \alpha}$) are related to each other by the Hermitian conjugation; all other creation  $\hat c^{\dag},~\hat b^{\dag},~\hat \gamma^\dag$ and annihilation $\hat c,~\hat b,~\hat \gamma$ superoperators (as well as their tilde-conjugated partners) are "canonically conjugated" to each other, i.e., for example,  $\hat c^{\dag}$ does not mean $(\hat c)^{\dagger}$ although $ \hat c\hat c^{\dag} \pm \hat c^{\dag}\hat  c =1$ ($\pm$ - bosons/fermions).

\subsection{Nonequilibrium quasiparticles}

Let us introduce nonequilibrium quasiparticle creation and annihilation superoperators, which are the basic building blocks for the NECI method.
We start by decomposing the Liouvillian as
\begin{equation}
  L = L^{(0)} + L',
\end{equation}
where $L^{(0)} = L^{(0)}_\mathrm{el}+ L^{(0)}_\mathrm{ph}$ is the Liouvillian for noninteracting electrons and phonons,
\begin{align}
  L^{(0)}_\mathrm{el}&=\varepsilon_0(\hat a^\dag\hat a - \widetilde a^\dag \widetilde a)+(\hat H_{SB} + \hat H_B -\mathrm{t.c.})-i\sum_{k\alpha}\Pi_{k\alpha},
                         \notag\\
 L^{(0)}_\mathrm{ph}&=\omega_0(\hat d^\dag \hat d - \widetilde d^\dag\widetilde d),
\end{align}
while  $L'$ represents the interaction between them
\begin{equation}
  L'=\kappa\{\hat a^\dag\hat a(\hat d^\dag + \hat d) - \mathrm{t.c.}\}
\end{equation}
Hereinafter, the notation 't.c' stands for tilde conjugated superoperators (see the tilde-conjugation rules in Appendix A).

Nonequilibrium quasiparticles are defined by diagonalizing the electronic part~of~$L^{(0)}$:
\begin{equation}
 L^{(0)}_\mathrm{el}=\sum_n(\Omega_n \hat c^\dag_n \hat c_n -\Omega^*_n \widetilde c^\dag_n \widetilde c_n),
\end{equation}
where $1\le n\le 2N+1$. Using the equation of motion method we find non-unitary (but canonical) Bogoliubov transformation which diagonalizes $ L^{(0)}_\mathrm{el}$:
\begin{align}\label{b_to_c}
  &\hat c_n = \psi_n \hat b + i\varphi_n\widetilde b^\dag + \sum_{k\alpha}(\psi_{n,k\alpha}\hat b^\dag_{k\alpha}+i\varphi_{n,k\alpha}\widetilde b^\dag_{k\alpha}),
  \notag\\
  &\hat c^\dag_n = \psi_n \hat b^\dag + \sum_{k\alpha}\psi_{n,k\alpha}\hat b^\dag_{k\alpha},~~
  \widetilde c_n  =  (\hat c_n)\,\widetilde{}~,~~ \widetilde c^\dag_n  =  (\hat c^\dag_n)\,\widetilde{}~,
\end{align}
where
\begin{align}
 & \hat b^\dag = \hat a^\dag - i\widetilde a,~~\hat b= \hat a,~~\widetilde b^\dag = (\hat b^\dag)\,\widetilde{},~~~ \widetilde b = (\hat b)\,\widetilde{},
  \notag\\
 & \hat b_{k\alpha}^\dag = \hat a^\dag_{k\alpha} - i\widetilde a_{k\alpha},~~\hat b_{k\alpha} =  (1-f_{k\alpha})\hat a_{k\alpha} - i f_{k\alpha} \widetilde a^\dag_{k\alpha},
  \notag\\
& \widetilde b_{k\alpha}^\dag = (\hat b_{k\alpha}^\dag)\,\widetilde{},~~~ \widetilde b_{k\alpha} = (\hat b_{k\alpha})\,\widetilde{},
\label{a_to_b}
\end{align}
Amplitudes $\psi$, $\varphi$  are the solution of the  following systems of equations
\begin{align}\label{sys1}
  &\varepsilon_0 \psi_{n}- \sum\limits_{k\alpha}t_{k\alpha}\psi_{n,b\alpha}=\Omega_n \psi_{n},
  \notag\\
  &E_{k\alpha}\psi_{n,k\alpha}-  t_{k\alpha}\psi_{n}=\Omega_n \psi_{n,k\alpha},
\end{align}
and
\begin{align}\label{sys2}
   &(\varepsilon_0-\Omega_n) \varphi_{n}- \sum_{k\alpha}t_{k\alpha} \varphi_{n,k\alpha}= \sum_{k\alpha} t_{k\alpha}f_{k\alpha} \psi_{n,k\alpha},
   \notag\\
  &(E^*_{k\alpha}-\Omega_n)\varphi_{n,k\alpha}-  t_{k\alpha}\varphi_{n}=-t_{k\alpha}f_{k\alpha} \psi_{n}.
\end{align}
with $E_{k\alpha}=\varepsilon_{k\alpha}+\Delta_{k\alpha}-i\gamma_{k\alpha}$.
By solving the eigenvalue problem ~\eqref{sys1} we also obtain the quasiparticle spectrum $\Omega_n$,~$-\Omega^*_n$.

Taking into account property~\eqref{bra_vac}, we see that $\bra{I}$ is a vacuum state for $\hat c^\dag_n$ and $\widetilde c^\dag_n$ superoperators,
 \begin{equation}\label{left_vacuum}
   \bra{I}\hat c^\dag_n = \bra{I}\widetilde c^\dag_n =0.
 \end{equation}

Although  superoperators $\hat c^\dag_n$ and $\hat c_n$ ($\widetilde c^\dag_n$ and $\widetilde c^\dag$)
are not Hermitian conjugate to each other, they obey the fermionic anticommutation relations. Particularly, from
 $\{\hat c^\dag_n, \hat c_{n}\}=1$ it follows that  $\psi$ amplitudes are normalized according to
\begin{equation}
  (\psi_n)^2 + \sum_{k\alpha}(\psi_{k\alpha})^2 = 1.
\end{equation}
Other useful relations between amplitudes $\psi,~\varphi$ can derived by making use the transformation inverse to~\eqref{b_to_c} (see relations~(A4) in our previous paper~\cite{dzhioev11a}) and computing the anticommutators.
For example
\begin{align}\label{useful_rel}
% & \{\hat b, \widetilde b\}=0,~ \Rightarrow~ \mathrm{Im}\sum_n \psi_n\varphi_n = 0;
% \notag  \\
 & \{\hat b, \widetilde b^\dag_{k \alpha}\}=0,~ \Rightarrow~ \sum_n (\varphi_{n,k\alpha}\psi_n - \psi^*_{n,k\alpha}\varphi^*_n) = 0;
 \notag  \\
 & \{\hat b, \hat b^\dag_{k \alpha}\}=0,~  \Rightarrow~ \sum_n \psi_{n,k\alpha}\psi_n = 0.
\end{align}

To compute the steady-state curent~\eqref{exp_value} we need to express the current superoperator $\hat J_\alpha$ in terms of
 quasiparticle superoperators.
By doing so and taking into account property~\eqref{left_vacuum}, we obtain the following expression for the steady-state current
\begin{align}\label{current_total}
  J_\alpha = \braket{I|\hat J_\alpha|\rho}
= -2\mathrm{Im}\sum_{m,k}t_{k\alpha}\psi_m\Bigl(\varphi_{m,k\alpha} +\sum_n\psi^*_{n,k\alpha}F_{mn}\Bigr),
\end{align}
where $F_{mn} = -i\braket{I|\widetilde c_n\hat c_m|\rho}$ is the two-quasiparticle amplitude. Note, that this is an \emph{exact} expression for the steady-state
current. The only problem is to find the unknown $F_{mn}$.

In the absence of electron-phonon interaction, i.e., when the nonequilibrium density matrix does not contain quasiparticle excitations,
the current through the central region is given by the first terms in~\eqref{current_total}. In what follows we will refer to this current as a free-field current~$J_\alpha^{(0)}$.
The correction to the free-field current, $\Delta J_\alpha$, is given by the second term involving $F_{mn}$.
In~\cite{dzhioev12}, it was shown how to compute $F_{mn}$ making use of the perturbation theory.
In the following subsections  we demonstrate how to find $F_{mn}$ within the NECI approach using
different reference states for the density matrix expansion, namely, free-field and coherent reference states.

\subsection{NECI expansion on free-field vacuum  as a reference state}

To write the configuration interaction expansion of the density matrix we need to define the reference state. In this paper, we consider two natural choices: a vacuum reference state (considered in the section - NECI expansion) and coherent state reference state (considered in section 3.3 , and denoted as NECI* expansion).
We define the free-field reference state as the steady-state density matrix in the absence of electron-phonon interaction, i.e.,
$\ket{\rho^{(0)} } = \ket{\rho^{(0)} }_\mathrm{el} \ket{\rho^{(0)} }_\mathrm{ph}$ and
\begin{equation}\label{free-field_rs}
  L^{(0)}\ket{\rho^{(0)} } = 0.
\end{equation}
The density matrix  $\ket{\rho^{(0)} }_\mathrm{el}$ is a vacuum state for nonequilibrium quasiparticles, i.e.,
\begin{equation}
  \hat c_n\ket{\rho^{(0)} }_\mathrm{el} =  \widetilde c_n\ket{\rho^{(0)} }_\mathrm{el}=0.
\end{equation}
To determine $\ket{\rho^{(0)} }_\mathrm{ph}$ we take into account the possibility that the phonon subsystem can contain a certain number of
thermally excited vibrational quanta. Let $N_\omega$ be the number of thermally excited phonons, i.e.,
\begin{equation}
 N_\omega=\bra{I} \hat d^\dag \hat d\ket{\rho^{(0)}}_\mathrm{ph}.
\end{equation}
It is convenient to perform a non-unitary canonical transformation and introduce new phonon operators
\begin{align}\label{d_to_gamma}
 &\hat \gamma =(1+N_\omega)\hat d - N_\omega \widetilde d^\dag,
 \notag\\
 &\hat \gamma^\dag = \hat d^\dag-\widetilde d
 \end{align}
and their tilde conjugated partners $\widetilde\gamma,~\widetilde\gamma^\dag$ such that $\ket{\rho^{(0)}}_\mathrm{ph}$ is the vacuum state
for $\hat\gamma, \widetilde\gamma$ superoperators,
\begin{equation}
 \hat\gamma \ket{\rho^{(0)}}_\mathrm{ph}=\widetilde\gamma \ket{\rho^{(0)}}_\mathrm{ph} = 0,
\end{equation}
while $\bra{I}$ is the vacuum for  $\hat\gamma^\dag, \widetilde\gamma^\dag$ superoperators (see property~\eqref{bra_vac} in Appendix). Now, the free-field Liouvillian
takes the form
\begin{equation}\label{L0_el}
  L^{(0)} = \sum_n(\Omega_n \hat c^\dag_n \hat c_n-\Omega^*_n \widetilde c^\dag_n \widetilde c_n) + \omega_0(\hat\gamma^\dag\hat\gamma - \widetilde\gamma^\dag\widetilde\gamma),
\end{equation}
and the vacuum state $\ket{\rho^{(0)}}$ for annihilation superoperators $\hat c_n,~\widetilde c_n,~\hat\gamma,~\widetilde\gamma$ is the free-field reference state obeying  condition~\eqref{free-field_rs}.
The free-field density matrix $\ket{\rho^{(0)} }$ is normalized according to $\braket{I|\rho^{(0)} }=1$.

Due to the electron-phonon interaction the exact steady-state density matrix contains multi-quasiparticle
and multi-phonon excitations above the free-field density matrix. We are looking for the exact steady-state density matrix in the form
\begin{equation}\label{rho_CI}
 \ket{\rho} = (1 + S)\ket{\rho^{(0)} },
\end{equation}
where the operator $S$ is given by the infinite power series of creation superoperators
\begin{align}\label{S_CI_full}
S = &\sum_{i=1}^\infty\alpha_i Q^\dag_i,~~~Q^\dag_i = \hat c_{m_1}^\dag\ldots\hat c^\dag_{m_k} \widetilde c_{n_1}^\dag\ldots \widetilde c^\dag_{n_l}(\hat\gamma^\dag)^p(\widetilde\gamma^\dag)^q.
\end{align}
Defined this way, the density matrix is normalized according to~$\langle I\ket{\rho}=1$, as
$\bra{I}$ is a vacuum for quasiparticle  and phonon creation superoperators.
Since the density matrix is tilde-invariant (see the definition in Appendix A), the superoperator $S$ obeys the property~$S=(S)\,\widetilde{}$~, i.e.
it remains the same if we complex conjugate all $\alpha_i$  and  replace the non-tilde superoperators by the tilde ones and vice versa.

We demand that the steady-state density matrix obeys the equation $L\ket{\rho}=0$, therefore
\begin{equation}\label{ful_CI}
  \Bigl\{SL' + [L^{(0)} + L',S]\Bigr\}\ket{\rho^{(0)} }= -L'\ket{\rho^{(0)} }.
\end{equation}
This superoperator equation is equivalent to the infinite inhomogeneous  system of linear equations for $\alpha_i$ amplitudes
in the expansion for nonequilibrium density matrix (\ref{S_CI_full}).
Any truncation in the expansion~\eqref{S_CI_full}   leads to approximate solution
of equation~\eqref{ful_CI}.

Here, we consider the simplest form of $S$ which allows us to calculate the correction for the free-field current. Namely,
\begin{equation}\label{S_simple}
  S = i\sum_{mn}(F_{mn}+Z_{mn}\hat\gamma^\dag + Z^*_{nm}\widetilde\gamma^\dag)\hat c^\dag_m \widetilde c^\dag_n + W(\hat\gamma^\dag+\widetilde\gamma^\dag),
\end{equation}
where $F_{nm}=F^*_{mn}$ and $W$ is a real number.
Inclusion of $W$ and $Z$ terms into the density matrix expansion  is necessary, since $\hat c^\dag_m\widetilde c^\dag_n\hat\gamma^\dag$,~~ $\hat\gamma^\dag$ configurations and their tilde-conjugate contribute
to the right-hand side of~\eqref{ful_CI} (see the expression for $L'$ below). If we neglect them, we observe a homogeneous linear system having a trivial solution. In this sense, these terms are
correlations inducing terms.

In order to  find equations for the amplitudes $F,~Z,~W$ we first express $L'$
in terms of nonequilibrium quasiparticle superoperators:
 \begin{align}\label{Lpr}
  L'=&\kappa\sum_{mn}\Bigl\{\bigl[L^{(1)}_{mn}\hat \gamma^\dag+L^{(2)}_{mn}\widetilde \gamma^\dag +L^{(3)}_{mn}(\hat\gamma+\widetilde\gamma)\bigr]\hat c^\dag_m \hat c_n -\mathrm{t.c.}\Bigr\}
  \notag\\
  -&i\kappa\sum_{mn}\bigl[L^{(4)}_{mn}\hat\gamma^\dag - (L^{(4)}_{nm})^*\widetilde\gamma^\dag+ L^{(5)}_{mn}(\hat\gamma+\widetilde\gamma)\bigr]\hat c^\dag_m\widetilde c^\dag_n
 \notag\\
  -&i\kappa\sum_{mn}L^{(6)}_{mn}(\hat\gamma^\dag-\widetilde\gamma^\dag)\hat c_m\widetilde c_n+ \kappa n^{(0)}(\hat\gamma^\dag-\widetilde\gamma^\dag).
\end{align}
Here the coefficients  $L^{(p)}_{mn}$ are
\begin{align}
  L^{(1)}_{mn}&=\bigl[(\psi_m-\varphi_m)+N_\omega\psi_m\bigr]\psi_n,
  \notag\\
  L^{(2)}_{mn}&=\bigl[\varphi_m+N_\omega\psi_m\bigr]\psi_n,~~L^{(3)}_{mn}=\psi_m\psi_n,
   \notag\\
  L^{(4)}_{mn}&=\bigl[(\psi_m-\varphi_m)\varphi_n^*+N_\omega(\psi_m\varphi^*_n-\varphi_m\psi_n^*)\bigr]
   \notag\\
  L^{(5)}_{mn}&=\psi_m\varphi_n^*-\varphi_m\psi^*_n,~~L^{(6)}_{mn}=\psi_m\psi_n^*,
\end{align}
and $n^{(0)}=\braket{I|\hat a^\dag \hat a|\rho^{(0)} }=\sum_n\psi_n\varphi_n$ is the free-field population of the electron level. Note, that after normal ordering $L'$ does not involve
terms containing annihilation superoperators only. Therefore, the condition $\bra{I}L=0$ is fulfilled.

Substituting $S$ given by~\eqref{S_simple} into~\eqref{ful_CI} and demanding equation~\eqref{ful_CI} to be fulfilled up to terms included into $S$ we derive
the system of linear equations for unknown amplitudes:

 \begin{align}\label{CI_system1}
   F_{mn}(\Omega_m  &- \Omega^*_n) + \kappa \sum_i L^{(3)}_{mi}[ Z_{i n}+ Z^*_{n i}]
    - \kappa\sum_i (L^{(3)}_{ni})^*[Z_{m i} +Z^*_{im} ]- 2\kappa L^{(5)}_{mn} W = 0
     \notag\\
   Z_{mn}(\Omega_m  &- \Omega_n^* + \omega_0) + \kappa n^{(0)} F_{mn}
+ \kappa \sum_i [L^{(1)}_{mi} F_{in} - (L^{(2)}_{ni})^* F_{mi}]   = \kappa L^{(4)}_{mn}
                            \notag \\
   W\omega_0 - &\kappa\sum_{mn} L^{(6)}_{mn}F_{mn} = -\kappa   n^{(0)}.
 \end{align}
It should be pointed out that the first and the last equations above are exact in the sense that the inclusion of other terms into $S$ does not modify these equations.

Before finishing this subsection it should be mentioned that Prosen has recently proposed in a series of papers\cite{Prosen2011_1,Prosen2011_2} a theoretical method for the exact solution of the nonequilibrium XXZ spin chain model. Prosen's method is based on the expression of nonequilibrium steady state matrix as a matrix product operator (MPO) with  infinite rank nearly diagonal matrices. If we perform Jordan-Wigner transformation to map Pauli matrices to fermion creation/annihilation operators, act by MPO $\hat{\rho}$ to left vacuum vector $  |I\rangle $, and  use our tilde substitution rules, then  MPO anzats of Prosen becomes physically equivalent to our NECI expansion~\eqref{rho_CI}. However, we would like to note  that
NECI expansion is more general than MPO anzats, because it is  not restricted to any specific Hamiltonian and applicable to nonequilibrium steady-state
in any many particle systems.

\subsection{Coherent reference state for density matrix expansion -- NECI*}

Within NECI method we have some flexibility in the choice of the reference state. Ideally, we would like to put as much
correlations as possible into the reference state while maintaining the possibility to define it as a vacuum for some quasiparticle annihilation operators.
 $L'$ given by~\eqref{Lpr} contains the term $\kappa n^{(0)}(\hat\gamma^\dag-\widetilde\gamma^\dag)$
which can be eliminated by the non-unitary canonical  transformation
\begin{equation}\label{gamma_to_xi}
  \hat\xi^\dag = \hat\gamma^\dag,~~~ \hat\xi = \hat\gamma + \kappa\frac{n^{(0)}}{\omega_0}
\end{equation}
and $\widetilde\xi^\dag=(\hat\xi^\dag)\,\widetilde{}$~, $\widetilde\xi=(\hat\xi)\,\widetilde{}$~.
The vacuum of the "shifted" phonon operators is the coherent state
\begin{equation}\label{rho*}
  \ket{\rho^{(*)} } = \exp\left\{-\kappa\frac{ n^{(0)}}{\omega_0}(\hat\gamma^\dag + \widetilde\gamma^\dag)\right\}\ket{\rho^{(0)} },
\end{equation}
normalized  according to $\bra{I}\rho^{(*)} \rangle=1$.
To distinguish from the NECI expansion on the free-field reference state, we denote the present method as NECI$^*$. The advantage of
NECI$^*$ approach is that it effectively includes multi-phonon excitation and de-excitation processes via the exponent~\eqref{rho*} in the coherent reference state.

The derivations are very similar to NECI presented in section 3.2. The only difference is that the Liouvillian and operator $S$ are expressed in terms of shifted phonon operators $\xi$:

\begin{equation}\label{L0}
  L^{(0)}= \sum_n(\Omega_n \hat c^\dag_n \hat c_n-\Omega^*_n \widetilde c^\dag_n \widetilde c_n) + \omega_0(\hat\xi^\dag\hat\xi - \widetilde\xi^\dag\widetilde\xi),
\end{equation}
 \begin{align}\label{Lpr2}
  L'=&- 2\kappa^2\frac{n^{(0)}}{\omega_0}\sum_{mn}\Bigl\{\bigl[L^{(3)}_{mn}\hat c^\dag_m\hat c_n -\mathrm{t.c.}\bigr] -iL^{(5)}_{mn}\hat c^\dag_m\widetilde c^\dag_n\Bigr\}
  \notag \\
 +&\kappa\sum_{mn}\Bigl\{\bigl[L^{(1)}_{mn}\hat\xi^\dag+L^{(2)}_{mn}\widetilde \xi^\dag +L^{(3)}_{mn}(\hat\xi+\widetilde\xi)\bigr]\hat c^\dag_m\hat c_n -\mathrm{t.c.}\Bigr\}
 \notag\\
  -&i\kappa\sum_{mn}\bigl[L^{(4)}_{mn}\hat\xi^\dag - (L^{(4)}_{nm})^*\widetilde\xi^\dag+ L^{(5)}_{mn}(\hat\xi+\widetilde\xi)\bigr]\hat c^\dag_m\widetilde c^\dag_n
\notag\\
  -&i\kappa\sum_{mn}L^{(6)}_{mn}(\hat\xi^\dag-\widetilde\xi^\dag)\hat c_m\widetilde c_n,
\end{align}
and
\begin{equation}\label{CI*_wf}
S = i\sum_{mn}(F_{mn} +Z_{mn}\hat\xi^\dag + Z^*_{nm}\widetilde\xi^\dag)\hat c^\dag_{m}\widetilde c^\dag_{n} + W(\hat\xi^\dag + \widetilde\xi^\dag).
\end{equation}

Demanding that  $(1+S)  \ket{\rho^{(*)} }$ is the steady state density matrix, we obtain
the following system of equations
   \begin{align}\label{CI*_system1}
   F_{mn}&(\Omega_m - \Omega^*_n) - \frac{2\kappa^2 n^{(0)}}{\omega_0}\sum_i [L^{(3)}_{mi}F_{in} - (L^{(3)}_{ni})^*F_{mi}]
    \notag\\
   &+ \kappa\sum_i L^{(3)}_{mi}[Z_{i n}+ Z_{n i}^*] - \kappa\sum_i (L^{(3)}_{ni})^*[Z_{m i} +Z^*_{im}]
  - 2\kappa L^{(5)}_{mn} W =  -\frac{2\kappa^2 n^{(0)}}{\omega_0}L^{(5)}_{mn}
   \notag\\
   Z_{mn}&(\Omega_m - \Omega_n^* + \omega_0) +\kappa \sum_i [L^{(1)}_{mi} F_{in} -  (L^{(2)}_{ni})^* F_{mi}]
   \notag\\
   &- \frac{2\kappa^2 n^{(0)}}{\omega_0}\sum_i [L^{(3)}_{mi}Z_{in} - (L^{(3)}_{ni})^*Z_{mi}]
 + \frac{2\kappa^2 n^{(0)}}{\omega_0} L^{(5)}_{mn}W =\kappa L^{(4)}_{mn}
                              \notag  \\
    W&\omega_0 -\kappa \sum_{mn} L^{(6)}_{mn}F_{mn} = 0.
 \end{align}
Solving this system we compute the two-quasiparticle amplitudes $F_{mn}$ and, hence, obtain the NECI$^*$ correction to the free-field current.

\subsection{Current-conserving properties of the theory}

The truncation of the density matrix expansion  introduces  approximations into the theory.
It is important to ensure that  observables computed with the truncated density matrix still satisfy conservation laws dictated by the symmetries of the underlying Hamiltonian.\cite{baym62,dahlen:164102, Kita2010}
For the electron transport problem it is particularly  important to demonstrate that the proposed configuration interaction theory preserves the particle number continuity equation
 in all orders of configurations included into the density matrix. To this end, we are going to prove that there is no artificial current leakage from the system introduced by the approximations and the current which enters the system from the left reservoir is exactly the same as the current which leaves the system to the right reservoir:
\begin{equation}\label{cur_conserv}
J_L  + J_R =0.
\end{equation}

Let us first  prove the current conservation in the free-field approximation. Indeed, the following equality is true
\begin{align}\label{current_total2}
  J^{(0)}_L&+J^{(0)}_R =  -2\mathrm{Im}\sum_{n,k\alpha}t_{k\alpha} \varphi_{n,k\alpha}\psi_n
  \notag\\
  &=-\mathrm{Im}\sum_{n,k\alpha}t_{k\alpha} (\varphi_{n,k\alpha}\psi_n - \psi_{n,k\alpha}\varphi_n)
  = \mathrm{Im}\sum_{n,k\alpha}t_{k\alpha}f_{k\alpha}\psi_{n,k\alpha}\psi_n = 0.
\end{align}
Here, we  have used relations~\eqref{useful_rel} as well as the first equations in~\eqref{sys1} and~\eqref{sys2}.

The solution of system~\eqref{CI_system1} (or~\eqref{CI*_system1})  provides us the two-quasiparticle amplitudes $F_{mn}$ and the NECI (NECI$^*$) correction $\Delta J_\alpha$ to the free-field current (see equation~\eqref{current_total}).
With the help of the first equation in ~\eqref{sys1} and using the property $F_{nm}=F_{mn}^*$ we find
\begin{align}\label{CI_cons}
\Delta J_L &+ \Delta J_R =  - 2\mathrm{Im}\sum_{mn,k\alpha}t_{k\alpha}\psi_{n,k\alpha}^*\psi_m F_{mn}
\notag\\
&=-2\mathrm{Im}\sum_{mn}\psi_m\psi^*_n(\varepsilon_0 - \Omega^*_n) F_{mn}
 = -\mathrm{Im}\sum_{mn}\psi_m\psi^*_n(\Omega_m - \Omega^*_n) F_{mn}.
\end{align}
Expressing $(\Omega_m - \Omega^*_n) F_{mn}$ from the first \emph{exact} equation in~\eqref{CI_system1} (or~\eqref{CI*_system1}) and taking into
account the  explicit expressions for $L^{(3)}_{mn}$, $L^{(5)}_{mn}$ we find that $\Delta J_L + \Delta J_R =0$. Note that this result does not depend on the particular choice of $S$, i.e. it
is valid for any truncated NECI (or NECI$^*$) expansion.

Thus, the presented configuration interaction  theory is current-conserving in all orders of nonequilibrium quasiparticle configurations included into the density matrix expansion.

\section{Numerical calculations}

In our numerical calculations we  assume that the applied voltage $V$ symmetrically shifts the on-site energies, i.e., $\epsilon_{L,R} = \pm 0.5V$.
Additionally, we assume that the electrodes are half filled, i.e., the corresponding left and right chemical potentials are positioned at $\epsilon_{L,R}$.
We choose the following parameters of the electrodes: $V_L=V_R=2.5$,  $t_L = t_R = 1.0$ and the temperature is $T=0.1$.
We set the applied voltage $V=1.0$.

In what follows, in the calculations based on the Lindblad master equation, we take a finite chain of $N$ atoms from each electrode for a buffer zone.
Therefore the energy spectrum of each buffer zone is given by\cite{peskin2010}
\begin{equation}\label{spectrum}
\varepsilon_{k\alpha} = \epsilon_\alpha + 2V_\alpha\cos\Bigl(\frac{\pi k}{N+1}\Bigr),~~k=1,\ldots,  N.
\end{equation}
and the tunneling matrix elements in~\eqref{t} are
\begin{equation}
  t_{k\alpha} = t_\alpha \sqrt{\frac{2}{N+1}}\sin\Bigl(\frac{\pi k}{N+1}\Bigr).
\end{equation}
The parameter $\gamma_{k\alpha}$ in Lindblad operators~\eqref{L_operators} is  taken to be equal the distance between neighbor energy levels in the
buffer zones, i.e., $\gamma_{k\alpha} = \varepsilon_{k\alpha} - \varepsilon_{k+1\alpha}$.

\begin{figure}[t!]
 \begin{centering}
\includegraphics[width=0.7\textwidth]{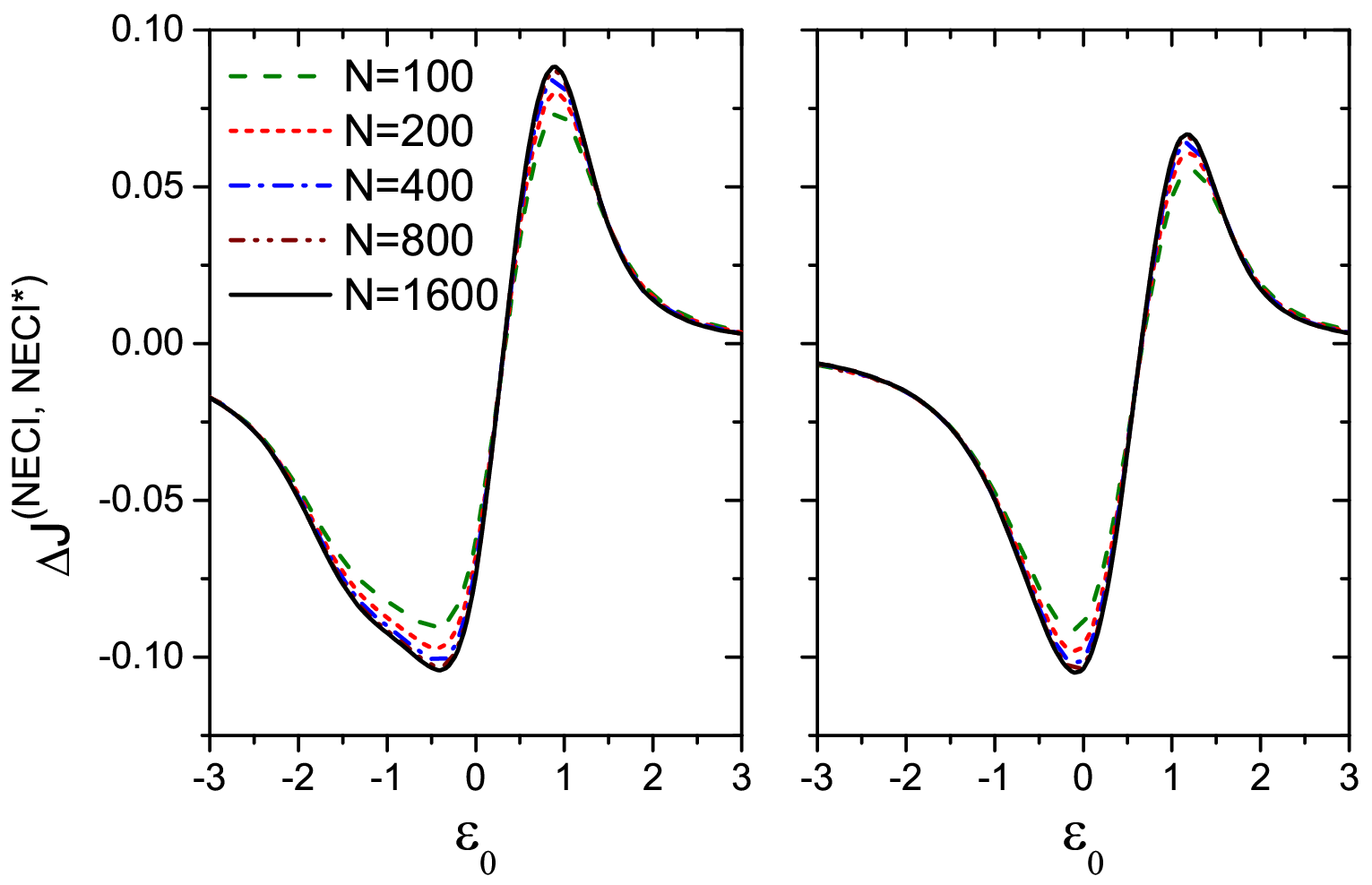}
\caption{NECI (left panel) and NECI$^*$ (right panel) corrections to the free-field current calculated for different values of $N$.
 The model parameters are  $\kappa = 1.0$, $\omega_0 = 1.0$, $N_\omega = 0$.}
 \label{limit}
 \end{centering}
\end{figure}

In our calculations we include $N=800$ sites into each buffer zone.
This size of the buffer zone has been proven to give the exact results for the  steady-state current calculated within the mean-field approximation and
the second-order perturbation theory.\cite{dzhioev11a,dzhioev11b} To justify this choice for NECI calculations we study
the convergence of our results with increasing the size of buffer zones.
In Fig.~\ref{limit} we present the results of such calculations for particular
values of model parameters.  Referring to the figure, we observe that
with increasing the size of buffer zones both $\Delta J^{(\mathrm{NECI})}$ and  $\Delta J^{(\mathrm{NECI*})}$ converge to some limit values and
for $N\ge 800$ both $\Delta J^{(\mathrm{NECI})}$ and  $\Delta J^{(\mathrm{NECI*})}$ are affected only marginally by increasing $N$.
We have performed the same calculations varying $\kappa,~\omega_0,~\mathrm{and}~N_\omega$ and found that this observation does not depend
on the particular choice  of model parameters. For this reason we take $N=800$ for the number of sites in each buffer zone. However, it should be pointed out,
 that rigorous results that the expectation values of operators reach their exact values when the size of the buffer zone increases
are applicable only to operators defined in the system subspace. In Ref.~\cite{Ajisaka2013} it is shown that the Lindblad master equation does not provide
a good description for the buffer zones itself, namely, the particle number distributions on the buffer zone differ slightly (by reasons
discussed in the paper) from expectations based on the Landauer formalism.

Let us say some words concerning numerical solution of the systems of equations~\eqref{CI_system1} and~\eqref{CI*_system1}. To be specific we consider the system obtained within NECI theory
(for the NECI$^*$ system we have used the same solution method).
The dimension of the system is of the order of $4M^2$, where $M=2N+1$. Although the systems is sparse, it contains about $12M^3$ nonzero matrix elements.
Assuming that each complex number requires 16 bytes of computer memory, we find that for $N=800$ the total system matrix needs about 700 gigabyte for storage.
A required memory size can be reduced drastically if we contract the indices by introducing the following linear combinations
\begin{equation}
  f_n = \sum_m F_{mn},~ z_n = \sum_m (Z_{mn} + Z^*_{nm})
\end{equation}
and their complex conjugate. Note, that the correction to the free-field current can be written as
\begin{equation}
  \Delta J_\alpha = - \sum_{n,k}t_{k\alpha}\psi^*_{n,k\alpha}f_n.
\end{equation}
Then, it is possible to rewrite the system~\eqref{CI_system1} as a system for $f_n$, $z_n$ and their complex conjugate. The obtained system contains $4M+1$ equations and the total
number of nonzero elements is about $6M^2$. To solve the system we have used standard routines from Intel MKL Fortran library.

\begin{figure*}[t]
 \begin{centering}
\includegraphics[width=1.0\textwidth]{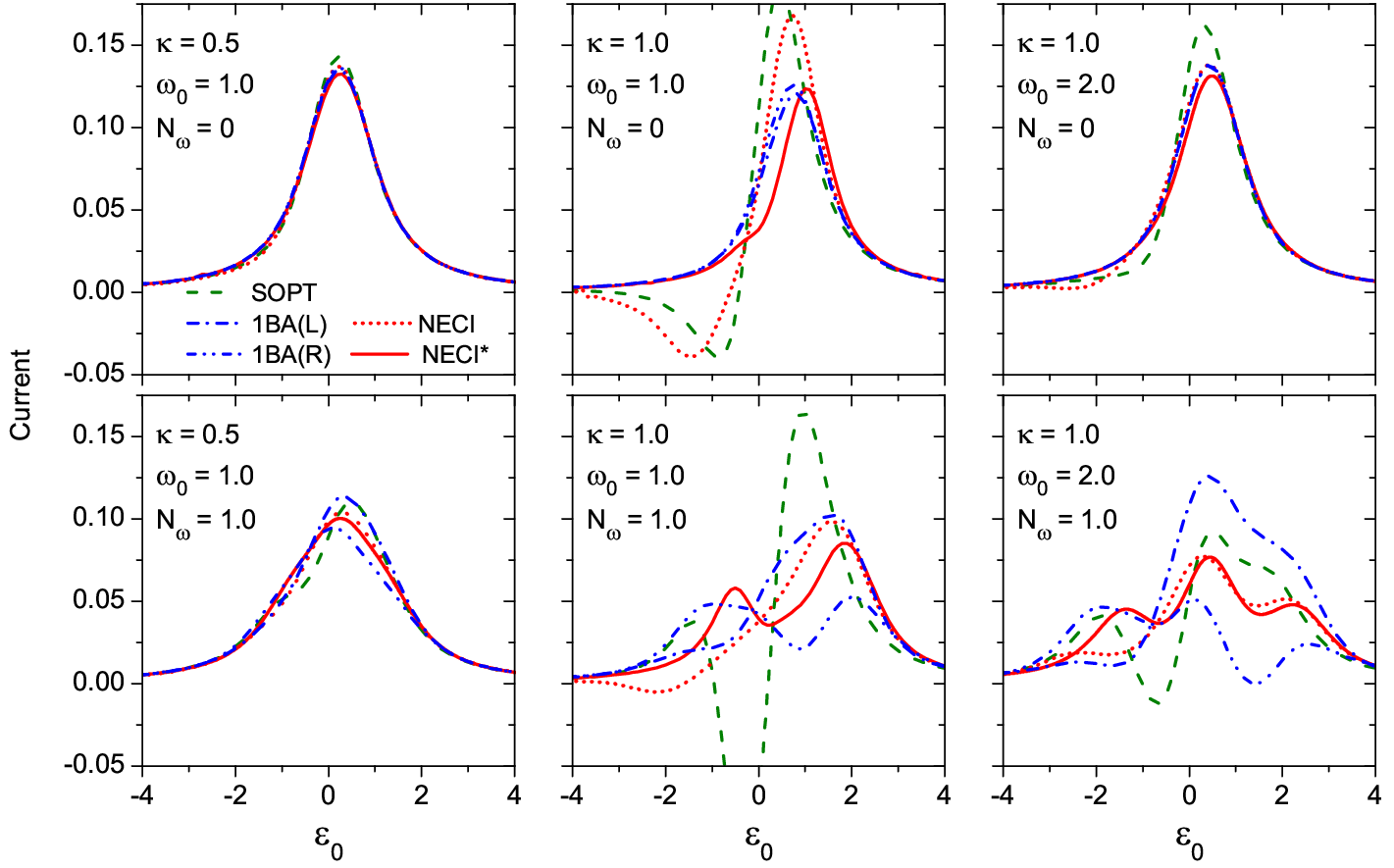}
\caption{Current through the central region calculated within different approaches as a function  of the level energy, $\varepsilon_0$.  }
 \label{fig2}
 \end{centering}
\end{figure*}

We also would like to compare the results obtained with the present approach with those calculated with the help of the nonequilibrium Green's function formalism.
For the net electric current passing through the $\alpha$ electrode to the central region we adopt the Meir-Wingreeen transport formula~\cite{meir92}
\begin{equation}\label{current_GF}
  J_\alpha = \int\frac{d\omega}{2\pi} \{\Sigma^<_{\alpha}(\omega)G^>(\omega) - \Sigma^>_\alpha(\omega)G^<(\omega)\}.
\end{equation}
Here $G^\gtrless$ are the greater and lesser Green's functions, and $\Sigma^\gtrless_\alpha$ are the greater and lesser electrode self-energies.
For the model of a  semi-infinite chain  the electrode self-energies can be written in an analytical form (see, for example, Ref.\cite{peskin2010}).
The  greater and lesser Green's functions can be obtained from the Keldysh equation~\cite{haug-jauho}. To solve this equation we apply
the first Born approximation (1BA) for the electron-phonon self-energy.~\cite{frederiksen2004}.
Although the 1BA does not generally guarantee current conservation, computationally it is not as demanding as the self-consistent Born approximation.
For weak electron-phonon coupling the second-order perturbation theory (SOPT) is a good approximation. In this approximation
the   greater and lesser Green's functions can be obtained from the 1BA ones by expanding them with respect to electron-phonon coupling constant
up to $\kappa^2$ terms.

In figure~\ref{fig2} we compare SOPT and 1BA  results for the electron current with the results
obtained within NECI and NECI$^*$ method  for different values of the central region parameters.
Since the first Born approximation does not guarantee the current conservation, we plot the 1BA current from the left electrode   to the central region as well as the 1BA current from the central region to the right electrode.
In Fig.~~\ref{fig2} the currents are shown as functions of the level energy, $\varepsilon_0$.
 We focus our attention to the case when $\varepsilon_0$ is situated
inside of the continuous spectrum of the leads. The latter is given by Eq.~\eqref{spectrum} at $N\to\infty$.
According to our numerical calculations, we get zero current through the system when goes to off-resonant when, i.e., regime $\varepsilon_0$ is far away from the leads energies.
We do not  observe any physically significant differences in the convergence between resonant  and off-resonant regimes.

We first  consider the case when $N_\omega=0$, i.e., there is no equilibrium thermally excited vibrational quanta. As evident from the figure
in this case the first Born approximation does not violate the current conservation significantly since the left and right currents are close to each other.
We see that for a weak coupling ($\kappa=0.5$) all approaches predict a similar $\varepsilon_0$ dependency of the current,  which reaches a maximum value
at $\varepsilon_0\approx \kappa^2/\omega_0$. The peak value of the SOPT and 1BA currents slightly exceeds the NECI and NECI$^*$ ones. When we increase the electron-phonon coupling
($\kappa=1.0$) both
SOPT and NECI currents become unphysical negative in the off-resonant regime when the electronic level $\varepsilon_0$ is below the electrode chemical potentials.
In the resonant regime SOPT and NECI currents significantly overestimate the results of  other approaches. In addition,
comparing NECI$^*$-based calculations with other calculations, we can see that NECI$^*$ approach gives the current peak position
at $\varepsilon_0\approx \kappa^2/\omega_0$.
If we now consider the case of larger phonon energy ($\omega_0 = 2.0$) we notice that negative SOPT and NECI currents in the off-resonant regime disappear
 and all approaches again demonstrate a similar $\varepsilon_0$ dependency of the current.

Now we consider results obtained for nonzero  number of   thermally excited equilibrium vibrational quanta. Namely, we assume that $N_\omega=1.0$.
In this case the first Born approximation clearly reveals its current nonconserving nature. This is illustrated in the lower panels of figure~\ref{fig2}
where we can see that the left and right 1BA currents can be very different  from each other. Nevertheless, in the weak electron-phonon coupling regime ($\kappa=0.5$), all approached predict
qualitatively similar dependence of the current upon the electronic level $\varepsilon_0$. Comparing the result for $\kappa=0.5$, $N_\omega=1.0$ with those for  $\kappa=0.5$, $N_\omega=0$
we notice that inclusion of  thermally excited equilibrium vibrational quanta into consideration produces slightly broadened current peak with a reduced amplitude. When the coupling
constant increases ($\kappa=1.0$) SOPT and NECI approaches again show unphysical negative current in the off-resonant regime when
$\varepsilon_0$ is below the electrode chemical potentials. Moreover, for $\kappa=1.0$ the electron-phonon coupling splits the NECI$^*$ current peak, while NECI approach gives only one peak.
For a larger phonon energy, $\omega_0=2.0$, NECI$^*$ predicts three pronounced peaks, while NECI gives only two peaks. In addition, SOPT and 1BA show unphysically negative current for certain values of $\varepsilon_0$.

From the above consideretion it is evident that the configuration interaction method  built on the coherent reference state is preferable when the electron-phonon coupling is large or when
there are thermally excited vibrational quanta in the system. This result is not surprising since,  contrary to NECI, NECI$^*$ method accounts for  multi-photon  excitations and de-excitation processes
which are important in these regimes.

Unfortunately, there is no exact result available for nonequilibrium Holstein model in the strong  leads-system coupling regime.
Therefore, we cannot access the deviation of the NECI$^*$  expansion from the exact solution at this moment
and we do not see how we can address this issue at the present stage of the theory development. In addition, we have found that for the considered choice of parameters the NECI$^*$ method
gives physically reasonable results up to $N_\omega \lesssim 20$ for $\kappa=0.5$ and  $N_\omega \lesssim 5$ for $\kappa=1.0$, but for larger values of $N_\omega$
the current becomes unphysical negative. This indicates that the simple truncation of NECI$^*$ expansion is not sufficient and
more complex multi-quasiparticle and multi-phonon terms should be included into the density matrix
expansion.

\section{Conclusions}

We developed nonequilibrium configuration interaction method, which formally gives the exact solution of out-of-equilibrium correlated many-body problem.
Our approach is based on a superoperator representation of the Lindblad master equation for the reduced density matrix of the embedded quantum system.
It was shown that the steady-state density matrix can be decomposed in Liouville-Fock space in terms of nonequilibrium multi-quasiparticle excitations above the reference vacuum
state. The amplitudes of these excitations provide a measure of the   many-body nonequilibrium correlations.

The theory was applied to study the inelastic electron transport  through the system with electron-phonon interaction.
To compute the current we used truncated expansion of the steady-state density matrix.
Two different reference states were considered: free-field vacuum and coherent state.
It was proved that both approximations are current conserving in all orders of the density matrix expansion.
The  current through the system was computed for different model parameters and compared with
the second-order perturbation theory and the first Born approximation results.  It was shown that the configuration interaction method based on the coherent
reference state is superior to the other approaches when the electron-phonon coupling is large or when
there are thermally excited vibrational quanta in the system.

The presented method can be readily applied for the electron transport calculations for the system with electron-electron correlations.
Extension of the method by inclusion higher order multi-quasiparticle terms in to the density matrix expansion is also possible.
Another route is to extend the method  to a dynamical nonequilibrium case by making the amplitudes in the density matrix expansion  time-dependent functions.
All these will be the subjects of our future investigations.

\appendix

\section{Overview of the superoperator formalism}

Here we briefly discuss the relevant to our paper details of  Liouville-Fock space superoperators, while
the comprehensive description  is given in~\cite{dzhioev12,schmutz78,prosen08}.
We suppose that the Fock space of the system under consideration is spanned by the complete orthonormal set of basis vectors $\ket{m}=\ket{m_1,m_2,\ldots}$ which are eigenvectors of the particle number operator:
\begin{equation}\label{compl}
  a^\dag_k a_k\ket{m}=m_k\ket{m},~~\sum_m \ket{m}\bra{m} = I,~~~\langle n | m \rangle = \delta_{nm}.
\end{equation}
The operators in the Fock space form themselves a linear  vector space called the Liouville-Fock space.
The set of vectors $\ket{mn}\equiv\ket{m}\bra{n}$
constitutes a  orthonormal basis in the Liouville-Fock space. Thus, every Fock space operator $A = \sum_{mn} A_{mn}\ket{m}\bra{n}$ can be considered as a
Liouville-Fock space ket-vector $\ket{A} = \sum_{mn}A_{mn}\ket{mn}$. The adjoint operator $A^\dag$ is represented by the bra-vector~$\bra{A}$.

The  scalar product in the Liouville-Fock space is defined as  $\braket{A_1|A_2} = \mathrm{Tr}(A_1^\dag A_2)$.
In particular, the scalar product of a vector $\ket{A}$ with $\bra{I}$
is equivalent to the trace operation in the Fock space, $\braket{I|A}=\mathrm{Tr}(A)$.

Now we introduce creation and annihilation superoperators.
Fermion creation and annihilation superoperators are defined as
\begin{align}\label{def_superop}
  \hat a_k\ket{mn} \equiv a_k \ket{m}\bra{n},~~  \widetilde a_k\ket{mn} \equiv i(-1)^\mu\ket{m}\bra{n} a^\dag_k
  \notag\\
 \hat a^\dag_k\ket{mn} \equiv a^\dag_k \ket{m}\bra{n} , ~~  \widetilde a^\dag_k\ket{mn} \equiv i(-1)^\mu\ket{m}\bra{n} a_k,
\end{align}
where $\mu = \sum_k(m_k + n_k)$. For  bosonic creation and annihilation superoperators we drop the phase factor $i(-1)^\mu$.
So defined superoperators satisfy the same (anti)commutation relations as their Fock space counterparts. Additionally,
the basis vectors of the Liouville-Fock space are super-particle number eigenvectors, i.e.,
\begin{equation}
\hat a^\dag_k \hat a_k\ket{mn} = m_k\ket{mn},~~ \widetilde a^\dag_k \widetilde a_k\ket{mn} = n_k\ket{mn}.
\end{equation}

For an operator $A = A(a^\dag, a)$  we formally define two superoperators
\begin{equation}\label{superA}
  \hat A = A(\hat a^\dag, \hat a),~~ \widetilde A = A^*(\widetilde a^\dag, \widetilde a)
\end{equation}
and refer to them as non-tilde and tilde superoperators, respectively.
The connection between non-tilde and tilde superoperators is given by the "tilde conjugation rules"
\begin{align}\label{TC_rules}
 &(c_1 \hat A_1 + c_2 \hat A_2)\widetilde{} = c_1^* \widetilde{A}_1 + c^*_2 \widetilde {A}_2,
 \notag\\
 &(\hat A_1\hat A_2)\widetilde{}=\widetilde{A}_1\widetilde{A}_2, ~~ (\widetilde A)\widetilde{} = \hat A,~~(\hat A^\dag)\widetilde{} = (\widetilde A)^\dag.
\end{align}

With the help of superoperators an arbitrary  Liouville-Fock space vector can be represented as
\begin{equation}\label{ketA}
  \ket{A} = \hat A\ket{I} = \sigma_A \widetilde A^\dag\ket{I}.
\end{equation}
Hereinafter, the phase $\sigma_A = -i$ if $A$ is a fermionic operator and $\sigma_A = +1$ if $A$ is a bosonic operator.
 We also define a Liouville-Fock space vector tilde conjugated to a given one, $\ket{A}\widetilde{}\equiv\widetilde A\ket{I}=\sigma_A^*\hat A^\dag\ket{I}$.
Therefore, if $A$ is a Hermitian bosonic  operator then $\ket{A}$ is tilde-invariant, i.e., $\ket{A}=\ket{A}\widetilde{}$.
The examples of tilde-invariant vectors are $\ket{I}$ and the density matrix $\ket{\rho}=\hat\rho\ket{I}$.

Considering the adjoint of~\eqref{ketA} and assuming that $\hat A=\hat a~\mathrm{or}~\hat a^\dag$  we find
\begin{equation}\label{bra_vac}
  \bra{I}(\hat a^\dag -  \sigma^*_a \widetilde a)=\bra{I}(\widetilde a^\dag -  \sigma_a \hat a)=0.
\end{equation}
This gives us the idea introduce superoperator $\hat b^\dag = \hat a^\dag -  \sigma^*_a \widetilde a$ and its tilde conjugate, $\widetilde b^\dag=\widetilde a^\dag -  \sigma_a \hat a$, which
annihilate the bra-vector $\bra{I}$.

For the product of operators the following relation is valid
\begin{equation}\label{A1A2}
  \ket{A_1A_2}=\hat A_1\ket{A_2} = \tau \widetilde A_2^\dag\ket{A_1},
\end{equation}
where $\tau=i$ if both $A_1$ and $A_2$ are fermionic and $\tau=\sigma_{A_2}$ otherwise. With the help of relations~\eqref{A1A2} the average of an operator $A$ in the state with the density
matrix $\rho$ can be calculated as the matrix element of the corresponding superoperator $\hat A$ sandwiched between $\bra{I}$ and $\ket{\rho}$:
\begin{equation}\label{average}
  \braket{A}=\mathrm{Tr}(A\rho)=\braket{I|A\rho}=\braket{I|\hat A|\rho}.
\end{equation}

Let us now consider the Lindblad master equation~\eqref{lindblad}. In the Liouville-Fock space this equation takes the form
\begin{equation}
  i\frac{d}{dt}\ket{\rho(t)} = \ket{H\rho(t)}-\ket{\rho(t)H} + i\ket{\Pi\rho(t)}.
\end{equation}
Applying raltions~\eqref{A1A2}, i.e., taking into account that $\ket{H\rho(t)}=\hat H\ket{\rho(t)}$, $\ket{\rho(t)H}=\widetilde H\ket{\rho(t)}$,
$\ket{a_{k\alpha}\rho(t)a^\dag_{k\alpha}}=\hat a_{k\alpha}\ket{\rho(t)a^\dag_{k\alpha}}=i\hat a_{k\alpha}\widetilde a_{k\alpha}\ket{\rho(t)}$, etc.,
we can rewrite this equation in the Schr\"{o}dineger-like form~\eqref{Schrodinger}. The time evolution of the density matrix is governed by
the non-Hermitian Liouville superoperator $L$. Two important properties of $L$ should be noted: 1. Since the density matrix is tilde invariant, then  $(L)\widetilde{}=-L$; 2.
Taking the time derivative of $\braket{I|\rho(t)}=1$ we find $\bra{I} L=0$, i.e., $\bra{I}$ is the left zero-eigenvalue eigenvector of $L$.

\section*{References}

\bibliographystyle{unsrt}

\end{document}